\documentstyle[aps,preprint,epsfig]{revtex}

\begin{document}

\input amssym.def
\input amssym.tex

\title{Transition-metal interactions in aluminum-rich intermetallics}

\author{Ibrahim Al-Lehyani$^{a,b}$ and Mike Widom$^{a}$}

\address{$^{a}$Department of Physics, Carnegie Mellon University,
Pittsburgh, Pennsylvania 15213}
\address{$^{b}$Department of Physics, King AbdulAziz University, Jeddah, Saudi Arabia}

\author{Yang Wang}

\address{Pittsburgh Supercomputer Center, Pittsburgh, Pennsylvania 15213}

\author{Nassrin Moghadam and G.~Malcolm Stocks}

\address{Oak Ridge National Laboratory, Oak Ridge, TN 37831-6114}

\author{John~A. Moriarty}

\address{Lawrence Livermore National Laboratory,
University of California, Livermore, California 94551}

\date{\today}

\maketitle

\begin{abstract}
The extension of the first-principles generalized pseudopotential theory 
(GPT) to transition-metal (TM) aluminides produces pair and many-body
interactions that allow efficient calculations of total energies.  In
aluminum-rich systems treated at the pair-potential level, one
practical limitation is a transition-metal over-binding that creates
an unrealistic TM-TM attraction at short separations in the absence
of balancing many-body contributions.  Even with this limitation, the 
GPT pair potentials have been used effectively in total-energy 
calculations for Al-TM systems with TM atoms at separations greater 
than 4~\AA.  An additional potential term may be added for systems with
shorter TM atom separations, formally folding repulsive contributions
of the three- and higher-body interactions into the pair potentials, 
resulting in structure-dependent TM-TM potentials.  Towards this end,
we have performed numerical {\it ab-initio} total-energy calculations 
using VASP (Vienna {\it Ab Initio} Simulation Package) for an Al-Co-Ni 
compound in a particular quasicrystalline approximant structure.  The 
results allow us to fit a short-ranged, many-body correction of the 
form $a(r_0/r)^{b}$ to the GPT pair potentials for Co-Co, Co-Ni, and 
Ni-Ni interactions.
\end{abstract}

\newpage

\section{Introduction}
\label{sec:intro}

Total-energy calculations are an important tool in theoretical
condensed-matter physics, giving insight into structures and mechanical
properties of solids~\cite{hafner87,pettifor95}.  Accurate calculations 
of total energy are notoriously difficult, however.  Theoretically, one 
must solve the Schr\"{o}dinger equation simultaneously for all electrons 
in the presence of fixed atomic nuclei.  Density-functional theory (DFT) 
\cite{hohenberg64,kohn65} simplifies this problem by reducing it to the
self-consistent solution of Schr\"{o}dinger's equation for a single
electron in a potential that depends upon the electron density.  Even
with this simplification, full {\it ab-initio} DFT electronic-structure 
methods are computationally demanding~\cite{AbInitio}, usually limited 
to systems of less than a hundred atoms, and may not yield immediate 
physical insight once an answer is obtained.

Instead, one may expand the total energy in terms of pair and many-body
interatomic potentials
\cite{hafner87,pettifor95,carlsson90,torrens72,interatomic88},
so that the total energy appears as an explicit function of atomic
separations.  Depending on the physical system under study and the type
of information sought, the expansion may often be truncated after a
small number of terms.  Such a truncated expansion trades off a degree
of accuracy in favor of computational simplicity and potentially
greater physical insight as compared with a full {\it ab-initio} 
electronic-structure approach.

Many metallic systems been studied using such quantum-based interatomic 
potentials, including aluminum and its alloys with both transition and 
non-transition metals 
\cite{moriartymcmahan82,krajcihafner92,phillipszou94,windischhafner94,mihalkzhu96,moriartywidom97}. 
These potentials are especially simple in the case of non-transition
metals.  There, the $d$-electron energy bands are either empty or else 
are completely filled and deeply buried below the Fermi energy level, 
allowing rapidly convergent expansions of the total energy and an accurate 
description in terms of only radial-force interactions~\cite{moriarty82}. 
The presence of partially filled $d$-bands at or near the Fermi level 
in transition metals complicates the analysis.  The occupied $d$-band 
electronic states are highly localized in the vicinity of the atoms, 
leading to directional or covalent bonding with a strong angular 
dependence.  Consequently, total-energy expansions will not converge 
as quickly as for non-transitions metals, and three- and higher-body 
angular-force interactions may contribute significantly
\cite{pettifor95,carlsson90,finnispaxton88,moriarty88}.

Moriarty~\cite{moriarty88} has developed a rigorous DFT treatment of 
interatomic potentials for transition metals (TMs) in the context of 
the generalized pseudopotential theory (GPT).  The treatment was later 
extended to binary and ternary alloys of aluminum with first row 
transition metals~\cite{moriartywidom97}.  These studies found that 
three- and four-body interactions could be important in determining 
thermodynamic and mechanical stability of structures with large TM 
concentrations.  The explicit treatment of $d$-electron interactions 
in the GPT produces a strong attractive interaction at unphysically 
short distances in the pair potentials, which is balanced by repulsive 
forces contained in the many-body interactions.  For specific 
structural environments, however, it is possible to directly modify the 
short-ranged part of the TM-TM pair potentials to remove this unphysical
attraction, so that a truncation of the total energy expansion at the
level of pair potentials will be more accurate when transition metal
atoms are near neighbors.  Here we wish to consider the construction
of such effective pair potentials for important transition-metal
aluminide systems.

One motivation for this study is the need for fast total-energy
calculations in Al-TM systems with short TM-TM separations to enable 
structural relaxation, and more generally, molecular-dynamics and 
Monte-Carlo simulations.  We focus our attention on Al-Co-Ni compounds 
in decagonal quasicrystalline structures~\cite{tsai89}.  The precise 
modifications required in the Co-Co, Co-Ni, and Ni-Ni pair potentials 
depend on the particular structure studied, but they should be at 
least approximately valid for many similar structures.  Furthermore, 
the modifications obtained may allow us to treat Al-Co-Cu and 
Al-Cu-Ni decagonal phases~\cite{otherdecag} because the Cu-Cu 
interactions do not appear to require modification~\cite{moriartywidom97}.  
Limited numbers of {\it ab-initio} electronic-structure calculations,
which effectively sum the pair and many-body total-energy contributions, 
are sufficient to determine the required modifications, and this is
the strategy that we follow here.

We intend to apply these potentials to predict the structures of
decagonal quasicrystals~\cite{allehyani_to}.  A great deal of
experimental data is available that identifies the positions of most
atoms and identifies the chemical identity of many of those.  However,
in order to determine the quasicrystal structures from X-ray
diffraction, one faces degenerate structures because elements near each
other in a row of the periodic table (such as Co, Ni and Cu) have
similar X-ray form factors.  A common approach to this problem is to
supplement the experimental data with total-energy calculations.  This
approach is well established in crystallography~\cite{XRAY}.

The effective pair potentials developed here can be applied to total 
energy calculations in quasicrystals and related structures with a great
reduction in computational times compared with the {\it ab initio}
electronic-structure calculations themselves. The time savings results 
from two general features of the potentials.  First, for a given atomic 
volume and composition the potentials may be
precalculated and then applied repeatedly with a simple lookup and 
interpolation. Second, to calculate the change in energy when a single 
atom is moved, only interactions affecting that atom are needed.  If 
the interactions are cut off at a finite spatial separation, the time 
required to calculate the change in total energy becomes independent of 
the number of atoms in the complete structure.  This is so-called 
order-$N$ scaling.  In contrast, {\it ab-initio} electronic-structure 
methods must recalculate the entire system when a single atom is moved, 
typically resulting in order-$N^3$ scaling.

In Sec.~\ref{sec:problem}, GPT interatomic potentials are briefly 
reviewed and the issues surrounding the truncation of the total energy 
expansion at the pair-potential level in Al-TM systems are discussed.  
Section ~\ref{sec:mod} gives details about the scheme we employ to 
determine the needed modifications to the TM-TM pair potentials.  In
Sec.~\ref{sec:results}, we discuss the results of our full {\it ab-initio} 
electronic-structure calculations and the modified TM-TM pair potentials 
developed using them.

\section{GPT Interatomic Potentials}
\label{sec:problem}

The generalized pseudopotential theory starts with a full {\it ab-initio}
DFT representation of the total energy in the standard local-density
approximation (LDA).  The usual small-core approximation is used to
separate the treatment of valence and core electrons and the electron-ion
interaction for the valence electrons is handled by means of optimized 
nonlocal pseudopotentials.  A mixed valence-wavefunction basis is employed 
allowing $sp$ states to be represented as superpositions of plane waves, 
while $d$ states are represented in terms of localized, atomic-like $d$ 
states.  The electron density and total energy are systematically expanded 
in terms of the resulting weak matrix elements in this basis: $sp$ 
pseudopotential matrix elements $W_{kk'}$, $sp$-$d$ hybridization matrix 
elements $\Delta_{kd}$ and $S_{kd}$, and $d$-$d$ tight-binding-like matrix 
elements $\Delta_{dd'}$ and $S_{dd'}$.  In real space, the total energy
may be cast in terms of well-defined interatomic potentials, which can be 
calculated as functionals of these matrix elements.  For a general 
multicomponent alloy, the GPT total-energy expansion takes the form
\begin{equation}
\label{eq:etot}
E_{\rm tot}({\vec{R}_{i}}) = NE_{\rm vol}(\Omega ,{\bf c}) +
\frac{1}{2}\sum_{\alpha \beta}\sum_{ij}\mbox{}^\prime \ 
v^{\alpha \beta}_{2}(R_{ij};\Omega ,{\bf c})+
\frac{1}{6}\sum_{\alpha \beta \gamma}\sum_{ijk}\mbox{}^\prime \ 
v^{\alpha \beta \gamma}_{3}(R_{ij},R_{jk},R_{ki};\Omega ,{\bf c})+\cdots \ ,
\end{equation}
where $\vec{R}$ is the set of all positions of $N$ ions in the metal, 
$E_{\rm vol}$ is a volume term that includes all collective and one-ion
contributions that are independent of structure, and $v^{\alpha \beta}_{2}$, 
$v^{\alpha \beta \gamma}_{3}, \ldots$ are the two-, three-, and higher 
many-ion interatomic potentials.  The primes on sums over ion positions 
exclude all self-interaction terms.  Indices $\alpha, \beta, \gamma, \ldots$ 
run over all chemical species, and indices $i, j, k, \ldots $ run over 
the individual ion positions.  The volume term and all of the interatomic 
potentials depend on the atomic volume $\Omega$ and a composition vector 
${\bf c}$, but are independent of structure.  The potentials are functions 
of the relative positions of small subsets of atoms, independent of the 
positions of all other atoms in the system.  The entire dependence on the 
structure comes analytically through the summations over ion positions.  
This makes these potentials transferable among different structures at 
fixed atomic volume and composition.  The full details of the first-principles
GPT for transition-metal systems are given in Refs.~\cite{moriarty88} and 
\cite{moriartywidom97}.  A simplified model version of the theory has 
also been developed~\cite{moriarty90}, using canonical $d$-bands to
obtain analytic representations of the multi-ion potentials. 

In general, the separation of the total energy into two- and higher-body 
terms is not entirely unique, since one can always add contributions to 
the pair potential $v^{\alpha \beta}_{2}$ provided one makes suitable 
subtractions from $v^{\alpha \beta \gamma}_{3}$ and/or higher-body 
potentials.  Within the GPT, the uniqueness of the potentials is 
established by ensuring that their desired properties of structure 
independence and full transferability are consistent with the matrix 
elements that define them.  In this regard, the total energy is normally
calculated to second order in the pseudopotential $W_{kk'}$, so that
$sp$ contributions enter only in the volume term and the pair potentials.  
The TM $d$-$d$ and $sp$-$d$ contributions to each potential, 
on the other hand, are carried to all orders in the matrix elements 
$\Delta_{dd'}$, etc.  Terms are allocated to pair- and many-body 
potentials according to how many distinct ionic positions explicitly 
enter.  Thus, for example, the TM pair potentials $v^{TM-TM}_2$ contain 
contributions that are even powers of $\Delta_{dd'}$ associated with 
repeated hopping of $d$ electrons back and forth between a pair of ions,  
with the leading term proportional to $\Delta_{dd'}\Delta_{d'd}$.  The 
three-ion TM potentials contain corresponding terms of third order 
proportional to $\Delta_{dd'}\Delta_{d'd''}\Delta_{d''d}$ and terms of 
fourth order proportional to $\Delta_{dd'}^2\Delta_{d'd''}^2$, as well 
as higher-order terms. The four-ion TM potentials start at fourth order 
in $\Delta_{dd'}$.

The tight-binding-like $d$-$d$ contributions to the TM potentials are 
modulated by $sp$-$d$ hybridization, $d$-state nonorthogonality, and 
other factors such as $d$-band filling, but nonetheless, they give 
valuable insight into the expected short-range behavior of the 
potentials.  In particular, one expects a strong attractive contribution 
to $v^{TM-TM}_2$ at short distances resulting from the second-order term 
in $\Delta_{dd'}$.  This term is attractive because it directly relates 
to the second moment of the $d$-band density of states and hence to the 
$d$-band width and the additional cohesion provided by partial $d$-band 
filling.  The attraction is strong at short distances because the matrix 
element for atoms separated by distance $r$ varies roughly as $r^{-5}$, 
the behavior obtained for pure canonical $d$ bands.  In addition, one 
expects this attractive contribution to be maximum near half-filling of 
the $d$ bands and to vanish for completely filled or empty $d$ bands.  
Thus the expected overbinding in $v^{TM-TM}_2$ will show a clear chemical 
dependence, with decreasing magnitudes for Co-Co, Co-Ni, and Ni-Ni 
interactions. 
 
For short-range TM interactions, repulsive contributions from higher-order 
terms in $\Delta_{dd'}$ will balance the attractive contribution of the 
second-order term in $v^{TM-TM}_2$, provided that the local concentration 
of TM atoms is sufficiently high.  In general, the detailed balance 
obtained involves the three-, four-, and possibly higher-ion potentials.  
Near half-filling of the $d$ bands, however, the repulsive contributions will 
be dominated by the fourth-order terms in $v^{TM-TM-TM}_3$.  This in turn
suggests a simple scheme to modify the TM pair potentials at short range 
to incorporate the balance directly, a scheme that we will develop in 
Sec.~\ref{sec:mod}. First, however, we examine the actual calculated GPT 
pair potentials for the Al-Co-Ni system of interest here.

\begin{figure}
\epsfig{figure=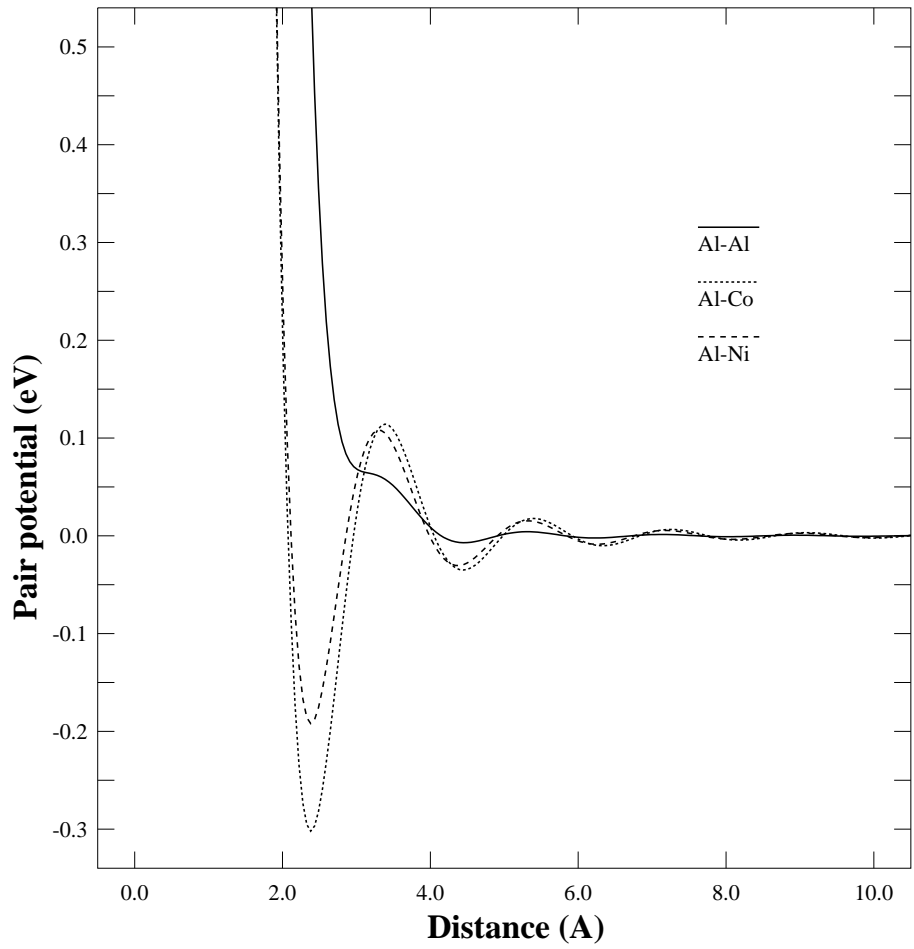}
\vspace{100pt}
\caption{GPT interatomic pair potentials for Al-Al, Al-Co and  Al-Ni.}
\label{fig:alal}
\end{figure}

\begin{figure}
\epsfig{figure=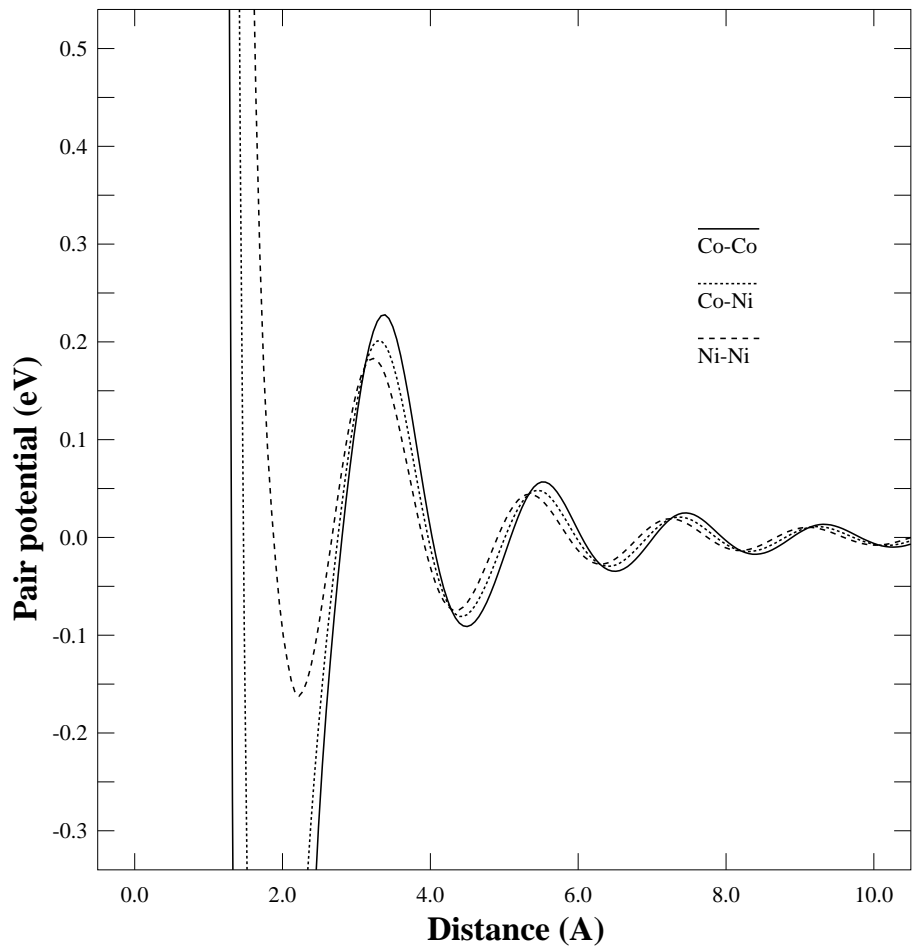}
\vspace{100pt}
\caption{GPT interatomic pair potentials for Co-Co, Co-Ni and Ni-Ni.}
\label{fig:tmtm}
\end{figure}

Figure~\ref{fig:alal} shows the Al-Al and Al-TM pair potentials for
Al-Co-Ni~\cite{moriartywidom97}.  These are calculated in the
aluminum-rich limit, but in practice they do not depend strongly on
composition.  The first minima of the Al-TM pair potentials occur near
2.3~\AA~with depths of about 0.2 eV (Al-Ni) and 0.3 eV (Al-Co). Rather 
than a potential minimum, the Al-Al potential exhibits a shoulder near 
3~\AA.  The TM-TM pair potentials are shown in Fig.~\ref{fig:tmtm}. 
As expected, the TM overbinding is most evident for Co.  The first
minimum in the Co-Co potential has a depth of 2.1 eV at 1.7~\AA.  The 
corresponding Ni-Ni potential depth of 0.1 eV at 2.2~\AA~ is not obviously 
problematic, but in the following we will find it requires some 
modification. In the present applications, the Co-Ni pair potential 
$v_2^{CoNi}$ is defined as an average of the Co-Co and Ni-Ni potentials,
\begin{equation}
\label{vconi}
v_2^{CoNi} \equiv (v_2^{CoCo}+v_2^{NiNi})/2 \ .
\end{equation}
This amounts to a perturbative expansion of $v_2^{\alpha\beta}$ in the
difference in atomic number $Z^{\alpha}-Z^{\beta}$. Clearly, $v_2^{CoNi}$ 
so-defined suffers overbinding due to the overbinding of $v_2^{CoCo}$.

We wish to devise effective pair potentials for Al-Co-Cu and Al-Cu-Ni 
as well as Al-Co-Ni. Previously, the Al-Cu potentials were found to be 
well behaved up to large Cu composition~\cite{moriartywidom97}, so no
modification of $v_2^{CuCu}$ is suggested.  Our modification to
$v_2^{CoCo}$ obtained for Al-Co-Ni compounds may be approximately
valid for these other compounds. We previously defined
\cite{moriartywidom97} $v_2^{CoCu}$ as equal to $v_2^{NiNi}$ because 
Ni lies between Co and Cu in the periodic table. Thus our modified 
Ni-Ni potential should serve as an approximate modified Co-Cu potential. 
For the modified Cu-Ni potential we may take $(v_2^{CuCu}+v_2^{NiNi})/2$, 
using the modified Ni-Ni potential. The Al-Co-Cu and Al-Co-Ni potentials 
so-obtained will, of course, still need to be validated using full 
{\it ab-initio} calculations.

\section{Modification of Pair Potentials}
\label{sec:mod}

As discussed above, the short-ranged attraction in the TM-TM pair 
potentials is balanced by repulsive terms contained in the three- 
and higher-body potentials.  If one chooses to truncate the GPT
expansion at the pair potential level, these repulsive many-body
contributions must be ``folded'' into effective pair potentials.
Formally, one may define an effective pair potential by averaging over
atomic positions, holding a single pair of ions fixed:
\begin{equation}
\label{veff}
v^{eff}_2 \equiv v^{\alpha \beta}_2 + <v^{\alpha \beta \gamma}_3> + 
<v^{\alpha \beta \gamma \delta}_4> + \cdots \ .
\end{equation}
Such potentials have been previously considered in the context of
the simplified model GPT~\cite{moriarty90} and canonical $d$ bands
for central transition metals.  There it was found that the four-body 
interaction oscillates with respect to angles between atoms, with a
nearly zero mean, so it does not contribute significantly to $v^{eff}_2$.  
The third-order contribution to $v^{\alpha \beta \gamma}_3$ also 
approximately averages away, but the fourth-order contribution to 
$v^{\alpha \beta \gamma}_3$ contributes strongly, yielding a short-ranged 
repulsive term proportional to $\Delta_{dd'}^2 \Delta_{d'd}^2 \sim
r^{-20}$ balancing against the attractive second-order term in 
$v^{\alpha \beta}_2$.

\begin{figure}
\epsfig{figure=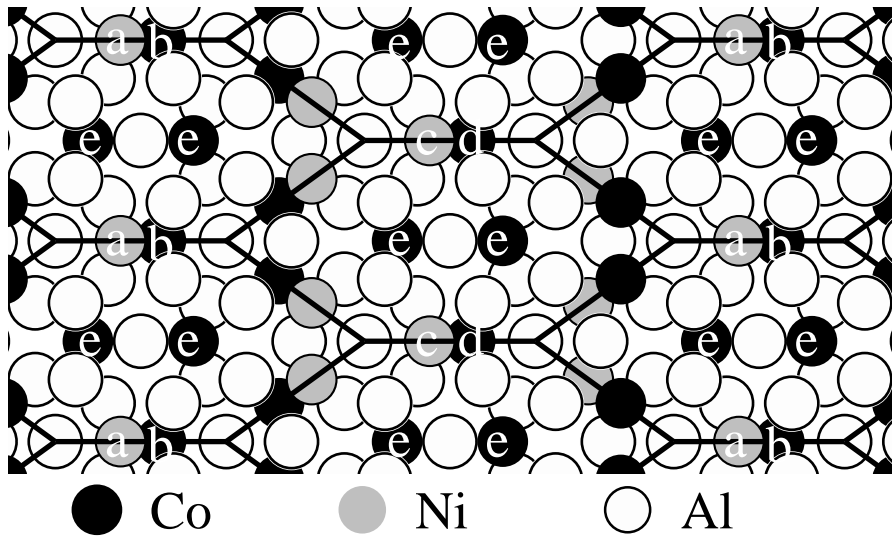}
\vspace{50pt}
\caption{The initial structure used in our calculations. Labeled atoms
participate in swaps.}
\label{fig:str}
\end{figure}

Inspired by the short-ranged repulsion found in Eq.~(\ref{veff}) 
and the power law variation of $\Delta_{dd'}$ within the model
GPT~\cite{moriarty90}, we propose to modify the full GPT pair
potentials $v_2^{\alpha\beta}$ by adding terms of the form
\begin{equation}
\label{Umod}
U^{\alpha\beta}(r)=a(r_{0}/r)^b \ ,
\end{equation}
where $a$ and $b$ are positive and depend upon the elements $\alpha$
and $\beta$ of the pair potential modified.  Our expectation, which is
confirmed below, is that $b$ is large in all cases, so that 
$U^{\alpha \beta}$ is indeed short-ranged.  In our applications, the 
quantity $r_{0}$ is taken as a common atomic separation in quasicrystals 
of 2.55 \AA. Then at a fixed atomic volume and composition the effective 
pair potential can be written as
\begin{equation}
\label{effpair}
V^{\alpha\beta}(r)=v_{2}^{\alpha\beta}(r)+U^{\alpha\beta}(r) \ .
\end{equation}
We determine the unknowns $a$ and $b$ by matching energies and forces 
obtained from full {\it ab initio} electronic-structure calculations 
on a quasicrystal approximant.  Cockayne and Widom
\cite{cockaynewidom98,widomallehyani00} previously suggested a structure
for decagonal Al-Co-Cu. An approximant of that structure is shown in
Fig.~\ref{fig:str} with Ni atoms replacing Cu. The orthorhombic unit
cell (a=23.3~\AA, b=7.57~\AA, c=4.09~\AA) contains 50 atoms
(Al$_{34}$Co$_{10}$Ni$_{6}$).  Most atoms occupy either
z=0.25 or z=0.75 layers.  Al atoms at the centers of hexagons occupy
the z=0.5 layer.  Two Co atoms occupy symmetric positions around these
central Al atoms. In Al-Co-Cu, alternation of Co and Cu on tile edges is
thought to be energetically advantageous~\cite{cockaynewidom98}.  We
find that alternation of Co and Ni shown in Fig.~\ref{fig:str} is
slightly {\it dis}advantageous in Al-Co-Ni.

To investigate TM bonding energetics, we alter the basic structure
shown in Fig.~\ref{fig:str} by swapping a Co atom on a horizontal tile
edge (atom b in Fig.~\ref{fig:str}) with the Ni atom on the other
horizontal tile edge (atom c). Focusing on near-neighbor interactions, 
we find this swap of atoms replaces four Co-Ni bonds with two 
Co-Co and two Ni-Ni bonds, all of length = 2.55~\AA. These numbers are
twice as large as is apparent by inspection of Fig.~\ref{fig:str}, the
extra factor of two coming from periodic boundary conditions in the
direction perpendicular to the plane.

Now consider the energy change evaluated using pair potentials. Atoms
b and c occupy nearly equivalent sites. An exact symmetry in the Al
atom positions guarantees that no bond involving an Al atom is
affected by the swap. We already noted the change in TM near-neighbor
interactions. At further neighbors, with separations of 4.6~\AA~ or
greater, we also find interchanges between Co-Co and Ni-Ni bonds for
pairs of Co-Ni bonds.  If the approximate form~(\ref{vconi}) of 
$v_2^{CoNi}$ as the average of $v_2^{CoCo}$ and $v_2^{NiNi}$ were valid, 
all changes in bonding would exactly cancel each other, resulting in a
vanishing energy change.  We presume that approximation~(\ref{vconi})
is more accurate at large separations than small separations. Thus
we attribute the entire energy change of the bc swap to near-neighbor
energy differences
\begin{equation}
\label{E1}
\Delta E_{1}=2V^{CoCo}+2V^{NiNi}-4V^{CoNi} \ ,
\end{equation}
where $V^{\alpha\beta}$ denotes the strength of the pair potential 
evaluated at the near-neighbor distance 2.55~\AA.

Next we swap one of the Co atoms inside the tiles (atom e) with one of
the Ni on a horizontal tile edge (atom a).  Two Co-Ni bonds are broken
and two Co-Co bonds are produced after this swap.  All other
interactions that are affected are Al-TM interactions, which we
presume to be described accurately by the GPT pair potentials.  This
swap energy can be written as:
\begin{equation}
\label{E2}
\Delta E_{2}=2V^{CoCo}-2V^{CoNi}+V^{AlTM} \ ,
\end{equation}
where $V^{AlTM}$ represents a calculable collection of interactions
between Al atoms and TM atoms at many separations.  $V^{AlTM}$ should
be described accurately by the unmodified GPT pair potentials.

Lastly, we replace the Co-Ni pair on one horizontal tile edge (atoms c 
and d) with Al atoms.  Then we swap one of the newly introduced Al
(at position c) with a Ni atom on the other horizontal tile edge (atom a).
This breaks two Co-Ni bonds.  All other interactions are either Al-TM
or Al-Al interactions, and again those are described well within the
GPT.  The energy change of this swap is
\begin{equation}
\label{E3}
\Delta E_{3}=-2V^{CoNi}+V^{AlTM}+V^{AlAl} \ ,
\end{equation}
where $V^{AlAl}$ and $V^{AlTM}$ represent collections of interactions
involving Al atoms that, as before, we presume to be accurately
calculable within the unmodified GPT.

Full {\it ab initio} values for the energy changes $\Delta E_1$, $\Delta
E_2$ and $\Delta E_3$ were calculated using VASP (Vienna {\it Ab Initio}
Simulation Package)~\cite{VASP}.  VASP calculates total energies within 
the local-density approximation using pseudopotentials to treat 
valence-core electron interactions.  We performed calculations using a 
4x4x4 k-space grid and also using a 4x4x8 k-space grid to observe the 
convergence as k-points are added.  All calculations were done using 
medium precision which is expected to be sufficient for our needs.  We 
iterate the self-consistent calculation until an accuracy of $10^{-6}$ 
eV is achieved.

By comparing the energy differences $\Delta E_1$, $\Delta E_2$ and
$\Delta E_3$ calculated by VASP with the same quantities calculated
with the unmodified GPT potentials, we can obtain the values of
$U^{\alpha\beta}$ evaluated at the near-neighbor separation
2.55~\AA. Specifically, when energy changes calculated by the 
unmodified GPT are subtracted from energy changes calculated by VASP, 
assuming that the contributions $V^{AlAl}$ and $V^{AlTM}$ are accurately 
calculated with the unmodified GPT, we find
\begin{eqnarray}
\label{simul}
\Delta E_1^{VASP}-\Delta E_1^{GPT} = 2U^{CoCo}+2U^{NiNi}-4U^{CoNi}\nonumber \\
\Delta E_2^{VASP}-\Delta E_2^{GPT} = 2U^{CoCo}-2U^{NiNi} \\
\Delta E_3^{VASP}-\Delta E_3^{GPT} = -2U^{CoNi} \ . \nonumber 
\end{eqnarray}
Since each correction $U^{\alpha\beta}(r)$ involves two unknowns, $a$
and $b$, Eq.~(\ref{simul}) consists of three equations in six unknowns. 
Additional information is obtained from the forces on atoms calculated 
by VASP. By examining the forces on the Co-Ni pair (atoms c and d) in 
Fig.~\ref{fig:str}, and on the Co-Co and Ni-Ni pairs created by the bc 
swap, we obtain three additional equations governing the derivatives of 
$U^{\alpha\beta}$ at the near-neighbor separation. This additional 
information allows closure of the equations and determination of the 
unknowns.

\section{Results}
\label{sec:results}

Table~\ref{tab:diff} shows the energy differences $\Delta E_{i}$ in
Eqs.~(\ref{E1})-(\ref{E3}) calculated using GPT pair potentials and VASP.  
Comparing the VASP data for the two grid sizes, we note that the signs 
and approximate magnitudes of $\Delta E_{i}$ are consistent with each 
other.  One immediate result from Table~\ref{tab:diff} is that
mixed Co-Ni bonds are disfavored over pure Co-Co and Ni-Ni bonds.  The
energy difference $\Delta E_{1}$ results from breaking four Co-Ni bonds
and producing two Co-Co and two Ni-Ni bonds.  $\Delta E_{1}$ calculated 
by VASP is negative, showing that the swap lowers the system energy.
This means that for Al-Co-Ni, similar TM atoms prefer to reside near
each other on the tile edges.  Cockayne and Widom found the opposite
for the case of Al-Cu-Co using mock ternary potentials
\cite{cockaynewidom98}, and this was confirmed later using a full 
{\it ab-initio} technique~\cite{widomallehyani00}.

Also concerning the calculated values of $\Delta E_1$, we see that
the averaged potential approximation~(\ref{vconi}) is fairly accurate. 
GPT yields $\Delta E_1 = 0$ because it employs this approximation here.
The small value of $\Delta E_1$ obtained by VASP confirms that this
approximation is not far off the mark.

\begin{figure}
\epsfig{figure=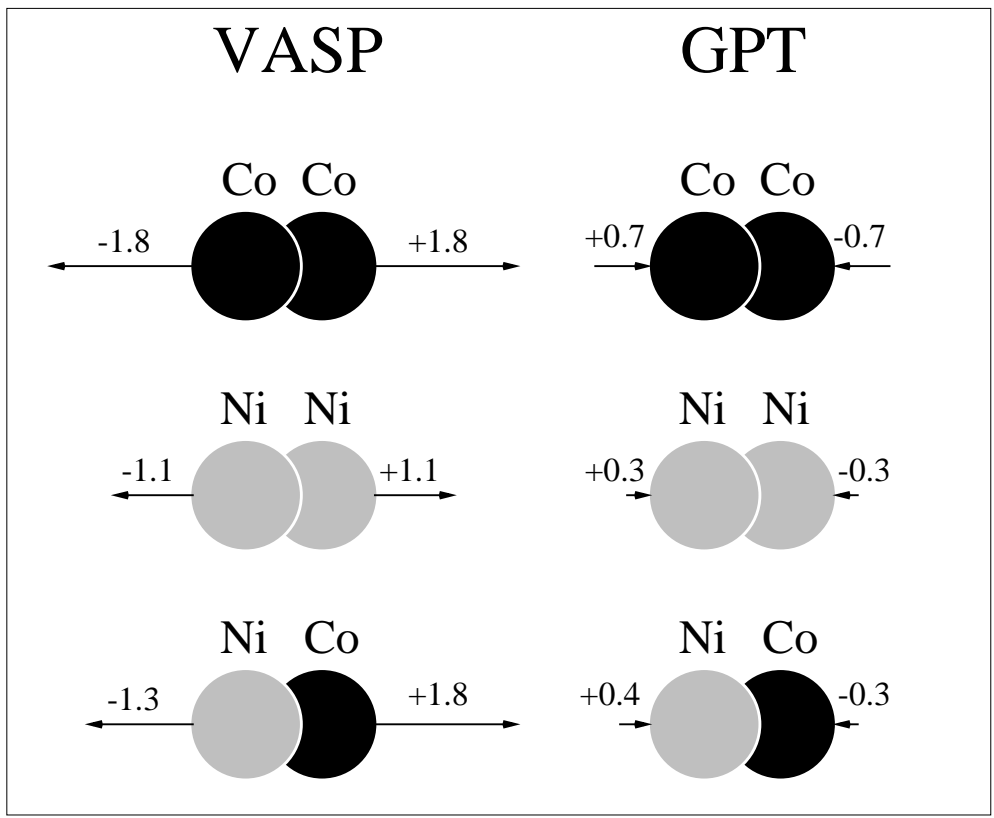}
\vspace{30pt}
\caption{Horizontal components of forces (in eV/\AA) on transition metal atom pairs calculated from the GPT and VASP.}
\label{fig:forces}
\end{figure}

Figure~\ref{fig:forces} shows the x-component of the total force on
certain TM atoms. Our (4x4x4) and (4x4x8) VASP calculations yield forces 
that agree to 0.06 eV/\AA~or better.  We examine the horizontal bonds ab and 
cd in Fig.~\ref{fig:str} in both the original and swapped configurations. 
As expected, at 2.55~\AA, GPT pair potentials predict attractive forces
between TM pairs while the actual forces obtained from VASP are
repulsive. The small force asymmetry on atoms in the Co-Ni pair is due
to the different ways Co and Ni atoms interact with their surrounding
environments. The difference between the forces calculated by our two
methods is greatest for Co-Co bonds and smallest for Ni-Ni bonds,
consistent with our expectation that overbinding is more severe for Co
than for Ni.

\begin{figure}
\epsfig{figure=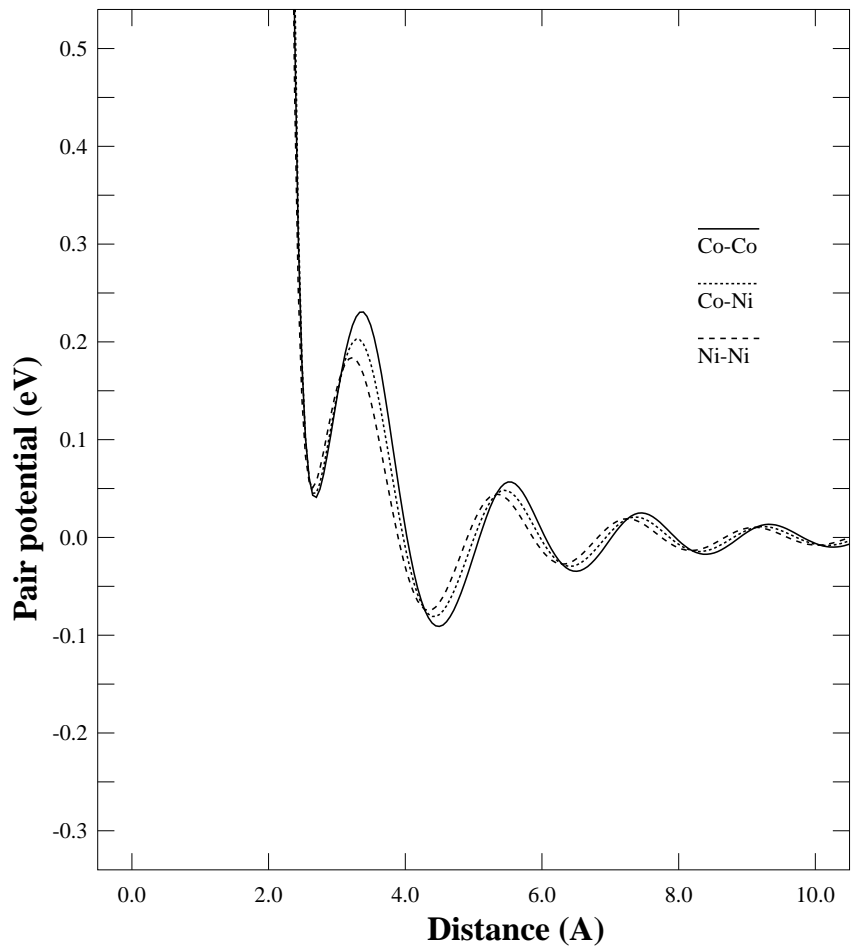}
\vspace{100pt}
\caption{Modified transition-metal GPT pair potentials using the parameters 
in Table~\ref{tab:para}.}
\label{fig:mod_tmtm}
\end{figure}

Calculated modifications to the GPT pair potentials are given in 
Table~\ref{tab:para}. Examining the magnitude of $U^{\alpha\beta}$ at
$r$=$r_0$=2.55~\AA~(i.e., the value of $a$), we note that $U^{NiNi}$ is 
smaller than $U^{CoCo}$, as is expected since Ni is closer to the end of 
the $3d$ transition series, with its $d$-bands almost full. It should be 
noted that $r$=2.55~\AA~is not the potential minimum. It is rather the 
nearest-neighbor distance at which the calculations were performed. The 
quantities $V^{\alpha\beta}$ and $F^{\alpha\beta}$ are, respectively, the 
energy and force calculated from the modified GPT potentials 
[Eq.(~\ref{effpair})] at the near-neighbor distance $r_0$.  The large 
values of $b$ we obtain show that our modifications 
of the GPT pair potentials fall off rapidly beyond the near-neighbor 
separation and confirm our expectations based on Eq. (3).  The modified 
potentials are illustrated in Fig.~\ref{fig:mod_tmtm}.  The (4x4x4) and 
(4x4x8) VASP calculations agree in positions of the potential minima to 
about 0.05~\AA~ and agree in the values at the minima to about 0.02 eV.

\section{Discussion}
\label{sec:dis}

The original GPT interatomic potentials were derived from first
principles without reference to specific structures.  Their
applicability to, and transferability among, a broad range of
structures was verified~\cite{moriartywidom97}. The pair potentials
alone apply to Al-rich structures in which TM atoms are well
separated, but fail due to an unphysical short-ranged TM
attraction. Our modification of the GPT potentials is restricted to TM
pair potentials. Since the correction $U^{\alpha\beta}(r)$ is
negligible beyond 3~\AA, our corrections only affect energies and
forces among neighboring TM atoms.

In idealized decagonal AlNiCo and AlCuCo quasicrystal models,
neighboring TM atoms always occur in specific atomic environments
consisting of zig-zag chains of TM atoms at 2.5~\AA~ spacing
surrounded by Al atoms at special positions~\cite{allehyani_to,cockaynewidom98,widomallehyani00}. A variety of quasicrystal and approximant structures differ in the arrangement of these chains in space, but share the same local
structure around the TM atoms. Our modified GPT potentials are
therefore strictly transferable within this class of quasicrystal
structures.

While the modified GPT potentials are not strictly transferable
outside this special class of structures, we do believe they are
qualitatively transferable. Comparing figures 2 and 5, we have
replaced obviously unphysical pair potentials with a set that appears
qualitatively realistic. Both original and modified pair potentials
are transferable among structures without TM neighbors. The modified
pair potentials provide a reasonable, though non-rigorous, extension
to structures with TM neighbors.

As a test of our modified pair potentials, we relaxed the basic
structure using both original and modified GPT pair potentials, and
using VASP. In all three runs we relaxed the structure until all
atomic forces were less than 0.001 eV/\AA. Under VASP relaxation
(4$\times$ 4$\times$4 k-space grid), Al atoms moved 0.19~\AA~ on average,
followed by Ni with an average displacement of 0.16~\AA~ and Co with
an average displacement of 0.15~\AA. Relaxation under the modified GPT
potentials produced a structure close to the VASP relaxed
structure.  The differences between modified GPT and VASP relaxed
positions are less than 0.07~\AA~ for every TM atom, with an average
difference of 0.05~\AA/TM atom. In contrast, relaxation under the
original GPT pair potentials yielded a maximum TM relaxed position
difference of 0.20~\AA~ and an average TM difference of 0.13~\AA.

As a result of our modifications, the TM atoms relax in the correct
directions and move approximately the correct distances. This is not
the case using the original GPT potentials. Since TM separations move
away from $r_0$ under relaxation, our proper behavior under relaxation
demonstrates a type of transferability of the potentials.

\acknowledgments
IA wishes to thank King Abdul Aziz University (Saudi Arabia) for
supporting his study and MW acknowledges support by the National
Science Foundation under grant DMR-9732567.  The work of the JAM was
performed under the auspices of the U. S. Department of Energy at the
Lawrence Livermore National Laboratory under contract number
W-7405-ENG-48. VASP calculations were performed at the Pittsburgh 
Supercomputer Center and at NERSC.

\begin{table}
\caption{Total-energy differences defined by Eqs.~(\ref{E1})-(\ref{E3})
as calculated by VASP and GPT. Units are eV/cell.}
\vspace{10pt}
\begin{tabular}{|l|r|r|r|}
 Energy  & GPT   & VASP (4x4x4) & VASP (4x4x8) \\
\hline
$\Delta E_{1} $       &  0.000	& -0.020 & -0.031 \\ 
$\Delta E_{2} $       &  0.116	&  0.298 &  0.279  \\  
$\Delta E_{3} $       & -0.945	& -1.384 & -1.419  \\    
\end{tabular}
\label{tab:diff}
\end{table}

\begin{center}
\begin{table}
\caption{Modifications for GPT transition-metal pair potentials, 
$U^{\alpha\beta}(r)= a(r_0/r)^b$ where $r_0$=2.55~\AA.  The quantities 
$V^{\alpha \beta}$ and $F^{\alpha \beta}$ are the energy and force 
calculated at $r$=$r_0$ from the modified GPT potential, Eq.~(\ref{effpair}).  
Units of $a$ and $V^{\alpha\beta}$ are eV while $b$ is dimensionless 
and $F^{\alpha\beta}$ has units of eV/\AA.}
\vspace{10pt}
\begin{tabular}{|l|c|c|c|c|}
$\alpha\beta$ &  $a$   & $b$ & $V^{\alpha\beta}$ & $F^{\alpha\beta}$  \\
\hline
CoCo        & 0.319 & 16.6 &0.0946   & 0.978  \\ 
CoNi        & 0.237 & 19.3 &0.0941   & 0.994 \\  
NiNi        & 0.140 & 21.3 &0.0779   & 0.674  \\    
\end{tabular}
\label{tab:para}
\end{table}
\end{center}

\newpage

\end{document}